\documentclass[prc,aps,groupedaddress,preprintnumbers,twocolumn]{revtex4}
\usepackage{graphicx,xcolor} %{graphicx,color}
\definecolor{Adam}{rgb}{0.9,0.1,0}

\begin{document}

\title{$0\nu\beta\beta$ nuclear matrix elements,
neutrino potentials and $\mathrm{SU}(4)$ symmetry}

\author{Fedor \v{S}imkovic}
\email[]{fedor.simkovic@fmph.uniba.sk}
\affiliation
        {\it  BLTP, JINR, 141980 Dubna, Moscow region, Russia
and Comenius University, Mlynsk\'a dolina F1, SK--842 48
Bratislava, Slovakia and IEAP CTU, 128--00 Prague, Czech Republic}
\author{Adam Smetana}
\email[]{adam.smetana@utef.cvut.cz}
\affiliation{\it Institute of Experimental and Applied Physics, CTU,
128--00 Prague, Czech Republic}
\author{Petr Vogel}
\email[]{pvogel@caltech.edu}
\affiliation
        {\it Kellogg Radiation Laboratory and Physics Department, 
Caltech, Pasadena, CA 91125, USA}

\date{\today}

\begin{abstract}
Intimate relation between the Gamow-Teller part of the matrix element $M^{0\nu}_\mathrm{GT}$ and the
$2\nu\beta\beta$ closure matrix element $M^{2\nu}_\mathrm{cl}$ is explained and explored. If the
corresponding radial dependence $C^{2\nu}_\mathrm{cl}(r)$ would be known,  $M^{0\nu}$ corresponding
to any mechanism responsible for the $0\nu\beta\beta$ decay can be obtained as a simple integral.
However, the $M^{2\nu}_\mathrm{cl}$ values sensitively depend on the properties of higher lying $1^+$
states in the intermediate odd-odd nuclei. We show that the $\beta^-$ and $\beta^+$ amplitudes
of such states typically have opposite relative signs, and their contributions reduce severally the  $M^{2\nu}_\mathrm{cl}$ 
values. Vanishing values of $M^{2\nu}_\mathrm{cl}$ are signs of a partial restoration of the spin-isospin
$\mathrm{SU}(4)$ symmetry. We suggest that demanding that $M^{2\nu}_\mathrm{cl}$ = 0 is a sensible way,
within the method of the Quasi-particle Random Phase Approximation (QRPA), of determining the amount of renormalization of isoscalar
particle-particle interaction strength $g^{T=0}_{pp}$.  Using such prescription, the matrix elements
$M^{0\nu}$ are evaluated; their values are not very different ($\le$ 20\%) from the usual QRPA
values when $g^{T=0}_{pp}$ is related to the known $2\nu\beta\beta$ half-lives.
\end{abstract}

%==============================================================================
\maketitle
%==============================================================================

%------------------------------------------------------------------------------
\section{Introduction}\label{S:intro} 
%------------------------------------------------------------------------------
Neutrinos are the only known elementary particles that may be Majorana fermions, i.e.,
identical with their antiparticles.
They are also very light, suggesting that the origin of their mass could be different from the origin
of mass of all other fermions that are much heavier and charged, supporting such hypothesis. 
Study of the neutrinoless double beta decay
($0\nu\beta\beta$), the transition among certain even-even nuclei when two neutrons bound in the
ground state are transformed into two bound protons and two electrons with nothing else emitted,
is the most straightforward test whether neutrino are indeed Majorana fermions.
Obviously, observing such decay would mean that the Lepton Number is not a conserved
quantity as required by the Standard Model. 

There is an intense worldwide effort to search for the $0\nu\beta\beta$ decay. No signal has been
observed so far, but  impressive half-life limits of more than $10^{25-26}$ years have been achieved in
several experiments on several target nuclei.  Larger, and even more sophisticated experiments are
developed and/or planned. Search for the $0\nu\beta\beta$ decay is at the forefront of
the present day nuclear and particle physics.

While observation of the $0\nu\beta\beta$ decay would constitute a proof that neutrinos are massive
Majorana fermions \cite{SV}, it is obviously desirable to be able to relate the observed half-life
to some `beyond the Standard Model' particle physics theory.  To do that, however, requires understanding
of the nuclear structure issues involved in the $(Z,A)_{\rm g.s.} \rightarrow (Z+2,A)_{\rm g.s.} + 2e^-$  
transition. The problem at hand is the evaluation of the corresponding nuclear matrix elements.
This is a long standing issue, with a plethora of papers devoted to this subject. Recent review
\cite{EM} summarizes the present status.

Here we explore in more detail the relation between the nuclear matrix elements
of the $0\nu\beta\beta$ decay and of the allowed and experimentally observed
$2\nu\beta\beta$ decay, treated however in the closure approximation. This is a continuation
and expansion of the earlier paper \cite{SHFV}. We concentrate primarily on the expression of these
matrix elements as functions of the relative distance $r$ between the two neutrons that are
transformed into the two protons in the $\beta\beta$ decay. Naturally, we keep in mind that
the closure approximation is not applicable for the $2\nu\beta\beta$ mode of the $\beta\beta$ decay. 

The paper is organized as follows. After this Introduction, in the next section the so-called neutrino
potentials are described, and their dependence on the distance $r$ between the decaying neutrons. 
Next, the two neutrino ($2\nu\beta\beta$) decay matrix elements in closure
approximation and their relation to the $0\nu\beta\beta$ decay matrix elements are discussed.
In the following section advantages of the $LS$ coupling scheme are described and symmetry
consideration are applied. In section \ref{S:SU} the $0\nu\beta\beta$ matrix elements, based
on previous considerations, are evaluated and their values are compared to the previously
published ones. The partial restoration of the spin-isospin symmetry $\mathrm{SU}(4)$ is also
discussed there. Finally, the Summary section concludes the paper.

Very generally, the observable $0\nu\beta\beta$ decay rate is expressed as a product of three
factors
\begin{equation}
\frac{1}{T_{1/2}} = G^{0\nu}(Z,E_0) (M^{0\nu})^2 \phi^2 ~,
\label{eq1}
\end{equation}
where $G^{0\nu}(Z,E_0)$ is the calculable phase space factor that 
in this case also includes all necessary 
fundamental constants, and that depends on the nuclear charge $Z$ and on the decay endpoint 
energy $E_0$. $M^{0\nu}$ is the nuclear matrix element that depends, among other things, on
the particle physics mechanism responsible for the the $0\nu\beta\beta$ decay,
as does the phase space factor $G^{0\nu}(Z,E_0)$. And
by $\phi$ we symbolically denote the corresponding particle physics parameter that we would like
to extract from experiment.  

For any mechanism responsible for the decay,
the matrix element  $M^{0\nu}$ consists of three parts, Fermi, Gamow-Teller and Tensor
\begin{equation}
M^{0\nu} = M^{0\nu}_{\mathrm{GT}} - \frac{M^{0\nu}_\mathrm{F}}{g_{\rm A}^2} + M^{0\nu}_{\rm T} ~,
\label{eq2}
\end{equation}
where $g_{\rm A}$ is the nucleon axial current coupling constant. And, in turn, the $\mathrm{GT}$ part, evaluated
in the closure approximation, is
\begin{equation}
M^{0\nu}_{\mathrm{GT}} = \langle f | \sum_{k,l} \vec{\sigma}_k \cdot \vec{\sigma}_l \tau^+_k \tau^+_l
H_{\mathrm{GT}}(r_{kl}, \bar{E}) | i \rangle ~.
\label{MGT}
\end{equation}
The Fermi part, again in closure, is given by an analogous formula
\begin{equation}
M^{0\nu}_{\mathrm{F}} = \langle f | \sum_{k,l}  \tau^+_k \tau^+_l H_\mathrm{F}(r_{kl}, \bar{E}) | i \rangle ~.
\label{MF}
\end{equation}
And the tensor part is
\begin{equation}
M^{0\nu}_{T} = \langle f | \sum_{k,l} [ 3 (\vec{\sigma}_k \cdot \vec{\hat{r}}_{kl}) 
(\vec{\sigma}_l \cdot \vec{\hat{r}}_{kl}) - \vec{\sigma}_k \cdot \vec{\sigma}_l ]\tau^+_k \tau^+_l
H_{\rm T}(r_{kl}, \bar{E}) | i \rangle ~.
\label{MT}
\end{equation}
Here $| i \rangle , | f \rangle$ are the ground state wave functions of the initial and final nuclei.
$H_{\mathrm{GT}}(r_{ij}, \bar{E})$,  $H_\mathrm{F}(r_{ij}, \bar{E})$ and $H_{\rm T}(r_{ij}, \bar{E})$ are the 
``neutrino potentials" that depend on the relative distance $r_{ij}$
of the two nucleons. The sum is over all nucleons in the nucleus. The dependence on the
average nuclear excitation energy $\bar{E}$ is usually quite weak. We discuss the validity
of the closure approximation for the $0\nu\beta\beta$ mode in the next section.

%------------------------------------------------------------------------------
\section{Neutrino potentials}\label{S:pot} 
%------------------------------------------------------------------------------

Neutrino potentials in eqs. (\ref{MGT}), (\ref{MF}) and (\ref{MT}) are typically defined as integrals 
over the momentum transfer $q$. They cannot be expressed by an analytic formula as functions of the 
internucleon distance $r_{ij}$. In the following we will concentrate on the ``standard" scenario, where
the $0\nu\beta\beta$ decay is associated with the exchange of light Majorana neutrinos. In that case
the particle parameter $\phi$ in eq. (\ref{eq1}) is the effective Majorana neutrino mass
\begin{equation}
m_{\beta\beta} = \left|\sum_{i=1}^{3} |U_{ei}|^2 e^{{\rm i}\alpha_i} m_i\right|,
\end{equation}
where $U_{ei}$ are the, generally complex, matrix elements of the first row of the PMNS neutrino mixing matrix with phases $\alpha_i$,
and $m_i $ are the masses of the corresponding mass eigenstates neutrinos. The present values
of the mixing angles and mass squared differences $\Delta m_{ij}^2$ are listed e.g. 
in the Review of Particle Properties \cite{PDG17}.  

For this mechanism, the dimensionless neutrino potential for the $K = \mathrm{GT}, \rm F$ and $\rm T$ parts is
\begin{eqnarray}
H_K (r_{12}, \bar{E}) &= & f^2_{\rm src}(r_{12}) \times\\
&&\frac{2}{\pi g_{\rm A}^2} {R} \int_0^{\infty}~ f_K(qr_{12})~
\frac{ h_K (q^2) q dq }
{q + \bar{E}} \,.\nonumber
\label{HGT}
\end{eqnarray}
here $R$ is the nuclear radius added to make the potential dimensionless.
The functions $f_{\mathrm{F},\mathrm{GT}}(qr_{12}) = j_0(qr_{12})$ and $f_{\rm T}(qr_{12})= - j_2(qr_{12})$
are spherical Bessel functions. The functions $h_K(q^2)$
are defined in \cite{anatomy}(see also \cite{SPVF}).
The potentials depend
rather weakly on average nuclear excitation energy $\bar{E}$.
The function $f_{\rm src}(r_{12})$ 
represents the effect of two-nucleon short range correlations. In the
following we use the $f_{\rm src}(r_{12})$ derived in \cite{Sim09}.
The phase space factors for this mechanism
are listed e.g. in \cite{KI12}.

However, the exchange of light Majorana neutrinos is not the only way $0\nu\beta\beta$ decay
can occur. Many particle physics models that contain so far unobserved new particles at the
$\sim\mathrm{TeV}$ mass scale also contain $\Delta L = 2$ higher dimension operators that could
lead to the $0\nu\beta\beta$ decay with a rate comparable to the rate associated with the
light Majorana neutrino exchange. 
These models also explain why neutrinos are so light. Moreover,
some of their predictions can be confirmed (or rejected) at the LHC or beyond.
Examples of these models are the Left-Right Symmetric Model or the R-parity Violating Supersymmetry. In them, heavy ($M \gg M_p$,
$M_p$ is the proton mass) particles are
exchanged between the two neutrons that are transformed into the two protons.
There is a large variety of neutrino potentials corresponding to such mechanisms of $0\nu\beta\beta$ decay.
A list of them, and of the corresponding phase space factors, can be found e.g. in ref. \cite{HN17}.
For a complete description of the $0\nu\beta\beta$ decay it would be, therefore, necessary to
evaluate $\sim20$ different nuclear matrix elements. We show below, how this task could
be substantially simplified.

 The matrix elements defined in the eqs. (\ref{MGT}), (\ref{MF}) and (\ref{MT}) are evaluated in the
 closure approximation. In that case only the wave functions of the initial and final ground states 
 are needed.  The validity of this approximation can be tested in the Quasi-particle Random Phase Approximation (QRPA), where the summation
 over the intermediate states is easily implemented as done in Ref. \cite{SHFV}. There it was
 shown that the closure approximation typically results in matrix elements that are at most 10\%
 smaller than those obtained by explicitly summing over the intermediate virtual states. The
 dependence on the assumed average energy $\bar{E}$ is weak; it makes little difference
 if $\bar{E}$ is varied between  0 and 12 MeV.  Similar conclusion was reached using the
 nuclear shell model (see Ref. \cite{SH} and references therein).

 Better insight into the structure of matrix elements can be gained by explicitly considering their
 dependence on the distance $r$ between the two neutrons that are transformed into two protons
 in the decay. Thus we define the function $C^{0\nu}_{\mathrm{GT}}(r)$ (and analogous ones for $M_\mathrm{F}$ and $M_{\rm T}$)
 as
 \begin{equation}
 C^{0\nu}_{\mathrm{GT}} (r)= 
 \langle f | \sum_{k,l} \vec{\sigma}_k \cdot \vec{\sigma}_l \tau^+_k \tau^+_l \delta(r - r_{kl})
H{_\mathrm{GT}}(r_{kl}, \bar{E}) | i \rangle ~.
\label{CGTdef}
\end{equation}
This function is, obviously, normalized as
\begin{equation}
M^{0\nu}_{\mathrm{GT}} = \int_0^{\infty} C^{0\nu}_{\mathrm{GT}}(r) dr ~.
\end{equation}
In other words, knowledge of $C^{0\nu}_{\mathrm{GT}}(r)$ makes the evaluation of $M^{0\nu}_{\mathrm{GT}}$ trivial.
The function $C(r)$ was first introduced in Ref. \cite{anatomy}.

\begin{figure}[htb]
%\centerline{\psfig{file=4thpaper_fig1.pdf,width=8.5cm}}
\includegraphics[width=\columnwidth]{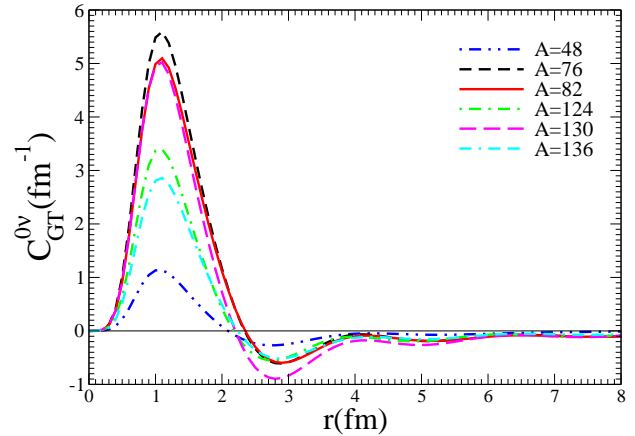}
\caption{Functions $C^{0\nu}_{\mathrm{GT}}(r)$  evaluated in the QRPA for a number of $0\nu\beta\beta$
candidate nuclei. }
\label{C0-qrpa}
\end{figure}

\begin{figure}[htb]
%\centerline{\psfig{file=4thpaper_fig1.pdf,width=8.5cm}}
\includegraphics[width=\columnwidth]{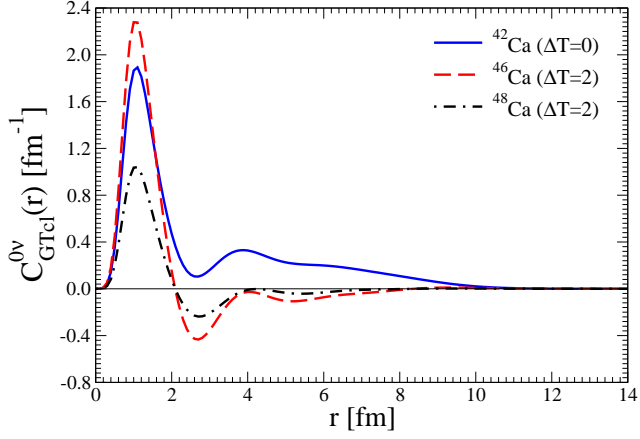}
\caption{Functions $C^{0\nu}_{\mathrm{GT}}(r)$  evaluated in the QRPA for several Ca isotopes.
$^{48}$Ca is a real $\beta\beta$ decay candidate. It decays into $^{48}$Ti and the isospin
$T$ changes in the decay by two units ($\Delta T = 2$). The other two Ca isotopes cannot
$\beta\beta$ decay; nevertheless the corresponding matrix elements can be evaluated. The
transition $^{42}$Ca $\rightarrow ^{42}$Ti connects mirror nuclei, the isospin does not change,
$\Delta T = 0$. }
\label{CaT=0,2}
\end{figure}

As one can see in Fig.\ref{C0-qrpa} the function $C^{0\nu}_{\mathrm{GT}}(r)$ consists primarily of a peak
with the maximum at 1.0-1.2 fm and a node at 2-2.5 fm. The negative tail past this node contributes
relatively little to the integral over $r$ and hence to the value of $M^{0\nu}_{\mathrm{GT}}$. The shape of the
function  $C^{0\nu}_{\mathrm{GT}}(r)$ is essentially the same for all $0\nu\beta\beta$ decay candidates.
The magnitude of the matrix element $M^{0\nu}_{\mathrm{GT}}$ is determined, essentially, by the value
of the peak maximum, which can be related, among other things, to the pairing properties of
the involved nuclei.

This characteristic behavior of the function $C^{0\nu}_{\mathrm{GT}}(r)$ repeats itself when it is evaluated
instead in the nuclear shell model; same peak, same node, little effect of the tail past the node
\cite{XM9}. The same function was also evaluated in \cite{P17} for the hypothetical decay
$^{10}$He $\rightarrow ^{10}$Be using the {\it ab initio} variational Monte-Carlo method. The
function $C^{0\nu}_{\mathrm{GT}}(r)$ has, again even in this case, qualitatively similar shape with a
similar peak and same node, but the negative tail appears to be somewhat more pronounced.
We might conclude that, at least qualitatively, the shape of $C^{0\nu}_{\mathrm{GT}}(r)$ is universal;
it does not depend on the method used to calculate it, even though the methods mentioned
here, QRPA, nuclear shell model, or the {\it ab initio} variational Monte-Carlo are vastly
different in the way the ground state wave functions $| i \rangle$ and $| f \rangle$ are 
evaluated.

In all $\beta\beta$ decay candidate nuclei the isospin $T$ of the initial nucleus is different,
by two units, from the isospin of the final nucleus; thus $\Delta T = 2$. To study theoretically 
nuclear matrix element evaluation it is not necessary to consider only the $\beta\beta$
transitions allowed by the energy conservation rules.   Thus, transitions within an isospin multiplet
($\Delta T =0$), such as $^{42}$Ca $\rightarrow ^{42}$Ti or $^6$He $\rightarrow ^6$Be can be,
and are, considered. The corresponding radial dependence $C^{0\nu}_{\mathrm{GT}}(r)$ is different in 
that case.
There is no node, the function remain positive over the whole $r$ range. For QRPA this is
illustrated in Fig. \ref{CaT=0,2}. Again, in the $\it ab~initio$ evaluation \cite{P17} for
the hypothetical transition
$^6$He $\rightarrow ^6$Be that feature is there as well, even though the shape of the curve
is rather different than for the  $^{42}$Ca case. The fact that the functions $C^{0\nu}_{\mathrm{GT}}(r)$
are quite different when $\Delta T =2$ and $\Delta T =0$ cases are considered, suggests
that it is not obvious whether the experience obtained from the latter cases in light nuclei
can be easily generalized to the decays of real $0\nu\beta\beta$ decay candidate nuclei
which are all $\Delta T =2$.

\begin{figure}[htb]
%\centerline{\psfig{file=4thpaper_fig1.pdf,width=8.5cm}}
\includegraphics[width=\columnwidth]{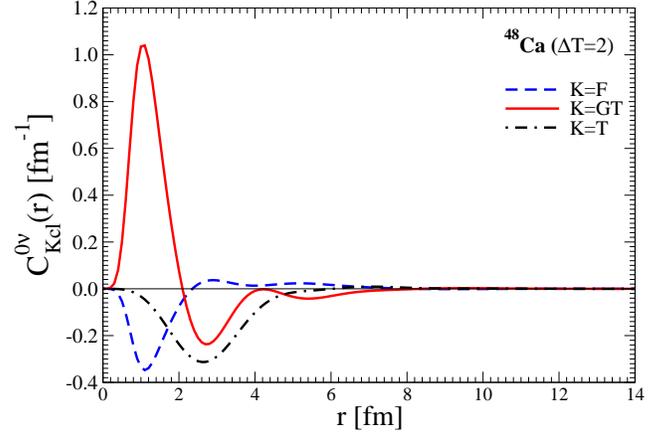}
\caption{Functions $C^{0\nu}_K(r)$  evaluated in the QRPA for the Gamow-Teller (GT), Fermi (F) and
Tensor (T) matrix elements for $^{48}$Ca $0\nu\beta\beta$ decay.}
\label{Cafgt}
\end{figure}

The radial functions $C^{0\nu}_\mathrm{F}(r)$ and $C^{0\nu}_\mathrm{T}(r)$ corresponding to the Fermi, eq. (\ref{MF}),
and Tensor, eq. (\ref{MT}), matrix elements are obtained in an analogous way. A typical example is
shown in Fig. \ref{Cafgt}. The function $C^{0\nu}_\mathrm{F}(r)$ has very similar shape as $C^{0\nu}_{\mathrm{GT}}(r)$,
but has opposite sign (see, however the sign in eq. (\ref{eq2})). The relation of $C^{0\nu}_\mathrm{F}(r)$
and $C^{0\nu}_{\mathrm{GT}}(r)$ will be discussed in detail in section \ref{S:LS}. Notice that the correlation
function $C^{0\nu}_\mathrm{T}(r)$ corresponding to the tensor matrix element does not share the properties of the main peak.

\section{ $2\nu\beta\beta$ matrix elements in closure approximation}
%\label{2nucl}

It would be clearly desirable to find a relation between the $0\nu\beta\beta$ matrix elements
and another quantity that does not depend on the unknown fundamental physics and that, in an
ideal case, is open to experiment. Here we wish to make a step in that direction.

If one would skip the neutrino potential $H{_K}(r_{ij}, \bar{E})$ in eq. (\ref{MGT}) the resulting
matrix element is  just the matrix element corresponding to the allowed $2\nu\beta\beta$ mode
of decay evaluated, however, in the closure approximation. The half-lives of $2\nu\beta\beta$ decay
have been experimentally determined for most candidate nuclei. They are related to the
matrix elements by
\begin{equation}
\frac{1}{T^{2\nu}_{1/2}} = G^{2\nu}(Z,E_0) (M^{2\nu})^2 ~,
\label{eq:2nu}
\end{equation}
where $G^{2\nu}(Z,E_0)$ is the calculable phase space factor that 
in this case includes all necessary 
fundamental constants, including the factor $g_{\rm A}^4$. The $2\nu\beta\beta$ matrix element, in turn, is
\begin{equation}
M^{2\nu} =  \sum_m \frac{ \langle f || \sigma \tau^+ || m \rangle 
\langle m ||  \sigma \tau^+ || i \rangle}{ E_m - (M_i + M_f)/2} ~,
\label{eq:2nme}
\end{equation} 
where the summation extends over all $1^+$ virtual intermediate states. The presence
of the energy denominators in eq. (\ref{eq:2nme}) is essential, it reduces the dependence on
the poorly known higher lying $1^+$ states.  Thus, if the $2\nu\beta\beta$ half-life is known
experimentally, the values of $M^{2\nu}$ can be extracted. (Actually, keeping in mind a possible
renormalization, i.e., quenching, of the $g_{\rm A}$ value in complex nuclei, the quantity
$g_{\rm A}^2 M^{2\nu}$ can be extracted from the experimental half-life value.)

\begin{figure}[htb]
%\centerline{\psfig{file=4thpaper_fig1.pdf,width=8.5cm}}
\includegraphics[width=\columnwidth]{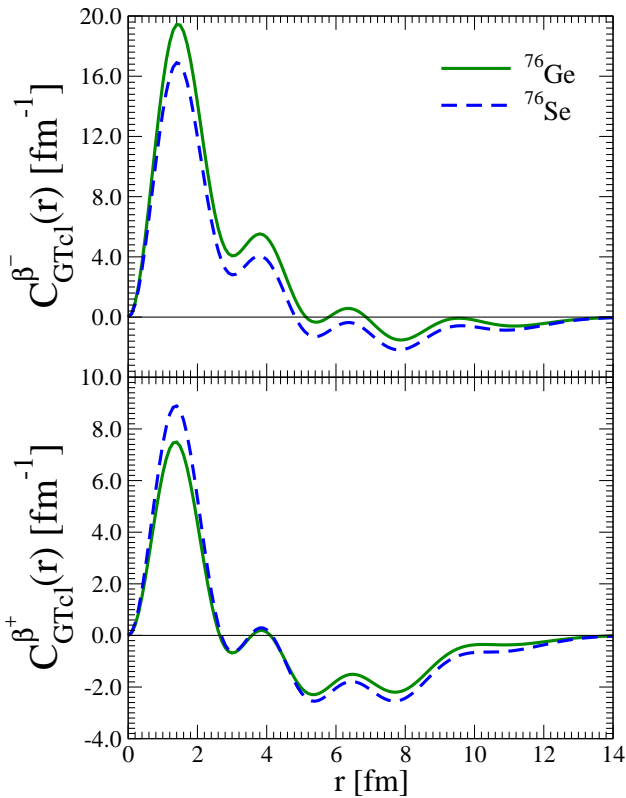}
\caption{Functions $C(r)$  corresponding to the total strengths $S(\beta^-)$ and $S(\beta^+)$
for the initial nucleus $^{76}$Ge and for the final nucleus $^{76}$Se.}
\label{fig:str}
\end{figure}

\begin{figure}[htb]
%\centerline{\psfig{file=4thpaper_fig1.pdf,width=8.5cm}}
\includegraphics[width=.50\textwidth]{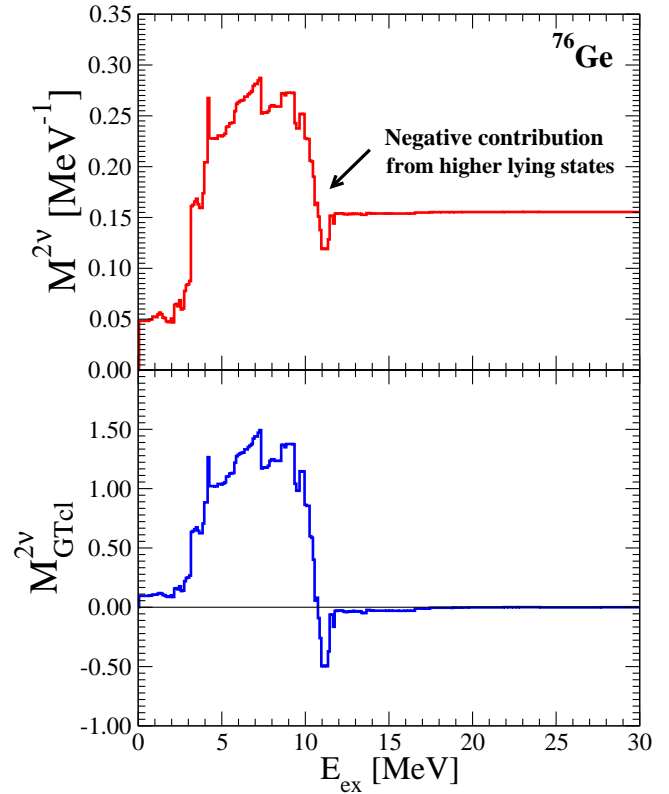}
\caption{Cumulative contributions to the $M^{2\nu}$ (\ref{eq:2nme})
 and $M^{2\nu}_{\mathrm{GTcl}}$ (\ref{2nucl}) as a function of the intermediate state excitation 
 energy. This is for the case of $^{76}$Ge.}
\label{fig:stairscl}
\end{figure}

Evaluation of the $2\nu\beta\beta$ closure matrix element
\begin{equation}
M^{2\nu}_{\mathrm{GTcl}} = \langle f | \sum_{k,l} \vec{\sigma}_k \cdot \vec{\sigma}_l \tau^+_k \tau^+_l| i \rangle 
=  \sum_m \langle f || \sigma \tau^+ || m \rangle  \langle m ||  \sigma \tau^+ || i \rangle ~
\label{2nucl}
\end{equation}
implicitly requires the knowledge of all $1^+$ intermediate states and the GT amplitudes connecting them
to the initial and final ground states. The expression (\ref{2nucl}) is a product of amplitudes corresponding
to the $\beta^-$ strength of the initial nucleus and the $\beta^+$ strength of the final one. The total 
strengths are connected by the Ikeda sum rule $S(\beta^-) - S(\beta^+) = 3(N-Z)$ which is automatically
fulfilled in the QRPA and in the Nuclear Shell Model when the model space involves both spin-orbit
partners of all single particle states. In Fig. \ref{fig:str} the radial dependence of these strengths, i.e., the $C(r)$ functions
corresponding to $\langle i |  \sum_{kl} \tau^+_k \tau^-_l \sigma_k \cdot \sigma_l | i \rangle $, i.e., the $S(\beta^-)$,
and $\langle f |  \sum_{kl} \tau^-_k \tau^+_l \sigma_k \cdot \sigma_l | f \rangle$, i.e., the $S(\beta^+)$,
are shown for the case of $^{76}$Ge and $^{76}$Se.
Note not only the different scales of the two panels, but also the substantial cancellation between
the $r \le 2.5$ fm and $r > 2.5$ fm in the $\beta^+$ case. The $S(\beta^+)$ strength is suppressed
because the $\beta^+$ operator connects states that belong to different isospin multiplets.

While the total strengths represent sums over positive contributions from all $1^+$ states in
the corresponding odd-odd nuclei, the $M^{2\nu}$ (\ref{eq:2nme}) and $M^{2\nu}_{\mathrm{GTcl}}$ (\ref{2nucl}) 
matrix elements both depend on the signs
of the two amplitudes involved in the product and thus have both positive and negative contributions. 
In fact, the calculations suggest that, as a function of the $1^+$ excitation energy, the
contributions are positive at first, but above 5 - 10 MeV negative contributions turn
the resulting values  of both $M^{2\nu}$ and $M^{2\nu}_{\mathrm{GTcl}}$ sharply down 
as illustrated in Fig.\ref{fig:stairscl}. 
That behavior seems to be again universal. 
Not only qualitatively similar curve are obtained in QRPA for essentially all
$\beta\beta$ decay candidate nuclei, but very similar plot was obtained for $^{48}$Ca
within the nuclear shell model \cite{Horoi07}.

In this context it is worthwhile to discuss the so-called single-state dominance 
(or low-lying states dominance) often invoked in the
analysis of the $2\nu\beta\beta$ decay \cite{SSD}. The `staircase' plot
for $M^{2\nu}$ evaluated within QRPA as seen in the upper
panel of Fig.\ref{fig:stairscl} have the drop at higher energies that
is not as steep as in the case of $M^{2\nu}_{\mathrm{GTcl}}$; its magnitude is reduced by the energy 
denominators. 

The contributions to $M^{2\nu}$ are positive at first, followed at 
energies $\ge$ 5 MeV by several negative ones. Due to this, the true value of $M^{2\nu}$ 
(0.14 MeV$^{-1}$ in the case of $^{76}$Ge) is
 reached twice as a function of the excitation energy, once at relatively low $E_\mathrm{ex}$
and then again at its asymptotic value. This is a typical situation encountered in most 
$2\nu\beta\beta$ decay candidate nuclei. In the charge exchange experiments, like e.g. 
in Ref. \cite{Grewe}, the GT strength exciting several low-lying $1^+$ states is determined
in both the $\beta^-$ and $\beta^+$ directions. Assuming that all contributions to the
$M^{2\nu}$ from these states are positive, one usually soon reaches a value that is close to the
experimental one. That is considered as indication of the validity of the  low-lying states dominance
hypothesis. 
The single (or low-lying) state dominance is also invoked in Refs. \cite{coello,ejiri12} where 
also a good
agreement with the experimental $M^{2\nu}$ matrix element was reached.
However, according to our evaluation, some more positive contributions to the
$M^{2\nu}$ in such a case are missed, as well as negative contributions from the higher lying
$1^+$ states. Thus, the low-lying states, while giving by themselves the correct (or almost correct)
value of $M^{2\nu}$, miss other contributions which, in particular, are decisively important for the
closure matrix element $M^{2\nu}_{\mathrm{GTcl}}$.

It would be clearly desirable to confirm, or reject, the behavior illustrated in Fig. \ref{fig:stairscl}.
In particular, to check that the $\beta^+$ amplitudes above $\sim$ 5 MeV are non-vanishing
and that their contribution to $M^{2\nu}$ is indeed negative. 

The single state dominance in the $2\nu\beta\beta$ decay can be tested by observing
the two- and single-electron spectra \cite{SSD1}, in particular at low electron energies. This
was done, for example, in the case of $^{82}$Se in Ref. \cite{SSD3}, indicating its validity.
Does it really mean that only low-lying intermediate states contribute to the $M^{2\nu}$ and
$M^{2\nu}_{\mathrm{GTcl}}$? As was shown in \cite{SSD2}, the deviation of the electron spectrum from
the standard form can be described by the Taylor expansion of the energy denominators when
the phase space factors are evaluated. The leading correction, called $\xi^{2\nu}_{31}$ there, 
contains the third power of the energy denominator in the expression analogous to (\ref{eq:2nme}).
Thus, the quantity $\xi^{2\nu}_{31}$ is dominated by the low lying states and insensitive
to the higher lying ones. The indication of single state dominance validity, like those in \cite{SSD3}, do not
mean that there are no higher lying contributions, and in particular a significant cancellations
in the  $M^{2\nu}_{\mathrm{GTcl}}$.

\begin{figure}[htb]
%\centerline{\psfig{file=4thpaper_fig1.pdf,width=8.5cm}}
\includegraphics[width=\columnwidth]{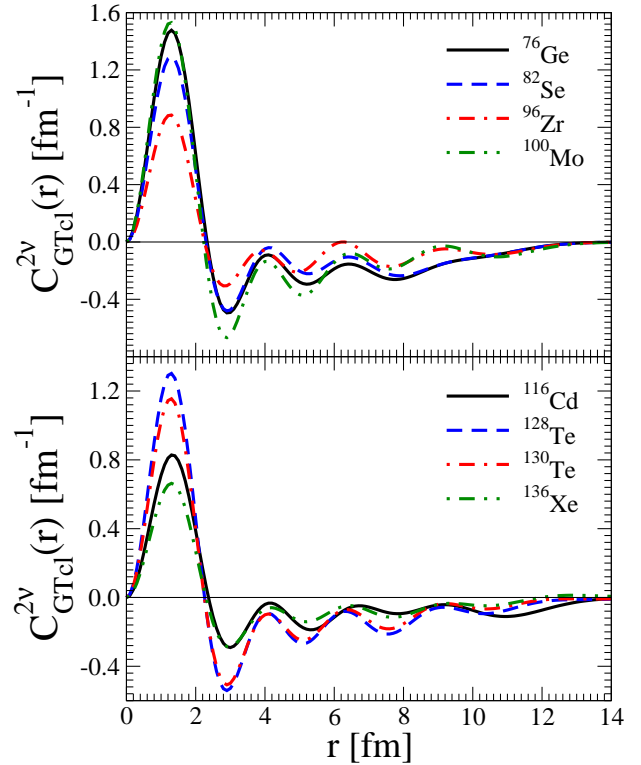}
\caption{Functions $C^{2\nu}_{\mathrm{GTcl}}(r)$ for several $\beta\beta$ candidate nuclei evaluated 
within the QRPA.}
\label{fig:cr_2ncl}
\end{figure}

The radial dependence $C^{2\nu}_{\mathrm{GTcl}}(r)$ corresponding to the $2\nu\beta\beta$ closure matrix element
(\ref{2nucl}) can be obtained, again, by inserting the Dirac $\delta$-function in between the
brackets. Note that while the closure matrix element (\ref{2nucl}) itself depends only on the
$1^+$ intermediate states, presence of the $\delta$-function means that all multipoles participate. 
In Fig.\ref{fig:cr_2ncl} we show the resulting radial function for a number of nuclei. The peak at
$r \le$ 2.5 fm is almost fully compensated by the negative tail at larger $r$ values. The actual value
of $M^{2\nu}_{\mathrm{GTcl}}$, while always small, depends sensitively on the input parameters (isovector
and isoscalar pairing coupling constants).

\begin{figure}[htb]
%\centerline{\psfig{file=4thpaper_fig1.pdf,width=8.5cm}}
  \includegraphics[width=\columnwidth]{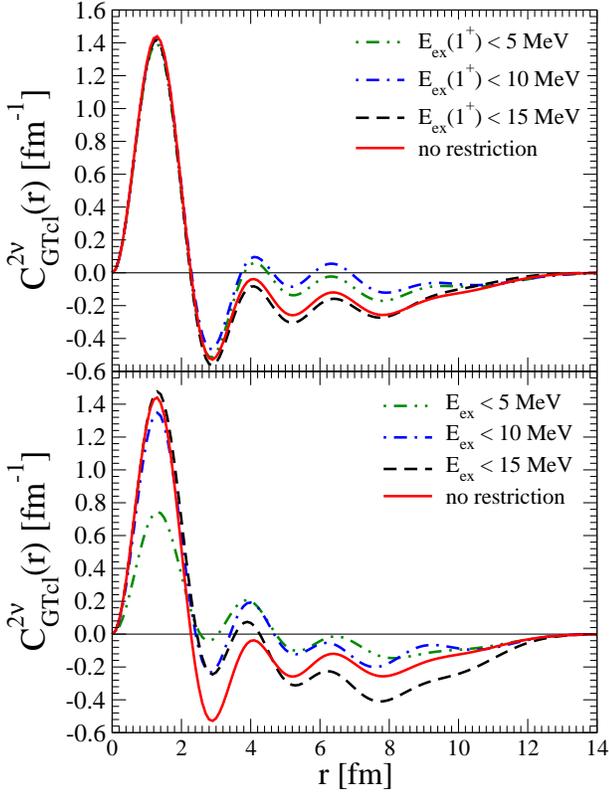}  
  \caption{Dependence of the  $C^{2\nu}_{\mathrm{GTcl}}(r)$ on the cut-off in the $1^+$ excitation energy
(upper panel) and all $J^\pi$ excitation energies (lower panel) evaluated for the $^{76}$Ge decay.}
\label{fig:ecut}
\end{figure}

It is important to add properly the contribution of all $J^\pi$ states when 
evaluating $M^{2\nu}_{\mathrm{GTcl}}$. In Fig.\ref{fig:ecut} 
we show how the corresponding $C^{2\nu}_{\mathrm{cl}}(r)$ depends on the possible energy
cut-off for all $J^\pi$ states (lower panel) and the $1^+$ states (upper panel).
The negative tail becomes deeper, and thus the magnitude of $M^{2\nu}_{\mathrm{GTcl}}$ becomes
smaller as more especially $1^+$ states are included.

Thus, when the $M^{2\nu}_{\mathrm{GTcl}}$ is evaluated in the shell 
model using incomplete oscillator shells, with missing spin-orbit partners, as done e.g. in Ref. \cite{Xav18}
for the $\beta\beta$ candidate nuclei (except $^{48}$Ca), the results might be uncertain.

\begin{figure}[htb]
%\centerline{\psfig{file=4thpaper_fig1.pdf,width=8.5cm}}
\includegraphics[width=\columnwidth]{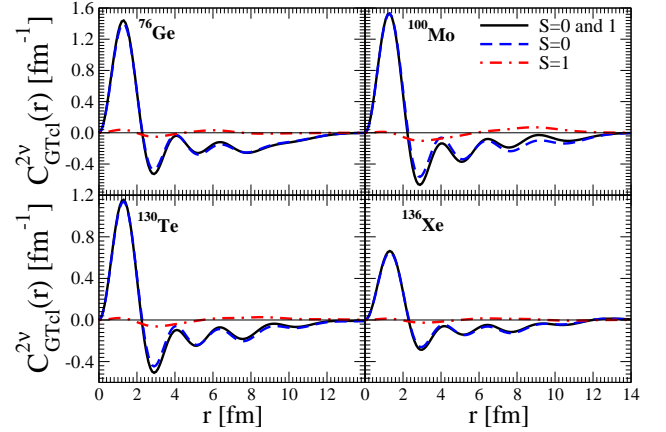}
\caption{Dependence of the  $C^{2\nu}_{\mathrm{GTcl}}(r)$ separated into the spin $S=0$ and $S=1$
components, shown for  several $\beta\beta$ decay candidate nuclei.}
\label{fig:s01}
\end{figure}

>From the way the functions $C^{0\nu}_{\mathrm{GT}}(r)$ and $C^{2\nu}_{\mathrm{GTcl}}(r)$ were constructed, it
immediate follows that they are related by
\begin{equation}
C^{0\nu}_{\mathrm{GT}}(r) = H_\mathrm{GT}(r,\bar{E}) \cdot C^{2\nu}_{\mathrm{GTcl}}(r) ~,
\label{eq:c02}
\end{equation}
as already pointed out in Ref. \cite{SHFV}. Therefore, if $C^{2\nu}_{\mathrm{GTcl}}(r)$
were known, the $C^{0\nu}_{\mathrm{GT}}(r)$ can be easily constructed and hence also
the $0\nu$ matrix element $M^{0\nu}_{\mathrm{GT}}$. The analogous procedure can be
followed, of course, also for $M^{0\nu}_\mathrm{F}$ and $M^{0\nu}_\mathrm{T}$. But eq. (\ref{eq:c02})
is much more general. Knowing $C^{0\nu}_{\mathrm{GT}}(r)$ or $C^{2\nu}_{\mathrm{GTcl}}(r)$ makes it
possible to evaluate the corresponding matrix element for any neutrino potential
$H\mathrm{GT}(r, \bar{E})$ like all of those listed in ref. \cite{HN17}. That represents, no doubt,
a significant practical simplification.

%----------------------------------------------------------
\section{Using the $LS$ coupling scheme}
\label{S:LS}
%-----------------------------------------------------------

>From the discussion above it is clear that the determination of the correct value of the
$2\nu$ closure matrix element $M^{2\nu}_{\mathrm{GTcl}}$ and its radial dependence function  $C^{2\nu}_{\mathrm{GTcl}}(r)$
is of primary importance. Insight into this issue can be gained by considering the $LS$
coupling scheme. 

Lets divide the  $M^{2\nu}_{\mathrm{GTcl}}$ and $M^{2\nu}_{\mathrm{Fcl}}$ into two parts, corresponding to the
$S = 0$ and $S = 1$, where $S$ is the spin of the two decaying neutrons (or spin of the
created protons) in their center-of mass system.  The corresponding expression is rather
complex so we leave it to the Appendix. Having the decomposition of the
$M^{2\nu}_{\mathrm{GTcl}}$ and its corresponding radial dependence $C^{2\nu}_{\mathrm{GTcl}}(r)$
into their spin  components, we can 
establish a relation between the GT and F parts.
\begin{eqnarray}
  M^{2\nu}_{\mathrm{Fcl}} &=& (\delta_{S1} + \delta_{S0}) \times  \langle s1,s2;S \parallel O^{2\nu}_{\mathrm{S}} \parallel s1,s2;S\rangle,
\nonumber  \\
M^{2\nu}_{\mathrm{GTcl}}&=&(\delta_{S1} -3 \delta_{S0})\times \langle s1,s2;S \parallel O^{2\nu}_{\mathrm{S}} \parallel s1,s2;S\rangle.
\nonumber\\
\label{eq:sepr}
\end{eqnarray}
Therefore, for the closure matrix elements 
\begin{eqnarray}
M^{2\nu}_{\mathrm{GT}, S=0} &=& -3 \times M^{2\nu}_{\mathrm{F}, S=0} \\ 
M^{2\nu}_{\mathrm{GT}, S=1} &=& M^{2\nu}_{\mathrm{F}, S=1} .
\label{eq:sep}
\end{eqnarray}
These are exact relations. The radial functions $C^{2\nu}_{\mathrm{Fcl,GTcl}}(r)(S)$ obey them as well.

Example of this separation are shown in Fig. \ref{fig:s01}. Clearly, the $S = 0$ component accounts
for essentially the whole $C^{2\nu}_{\mathrm{GTcl}}(r)$ function; the $S = 1$ component is negligible. 
Note that the standard like nucleon pairing 
supports the dominance of the $S = 0$ component.

Isospin is a good quantum number in nuclei, $T = (N - Z)/2$ in the ground states; the admixtures of
higher values of $T$ is negligible for our purposes. From this it immediately follows that
$M^{2\nu}_{\mathrm{Fcl}} = 0$. That relation is obeyed automatically in the nuclear shell model where
isospin is a good quantum number by construction. In QRPA, however, the isospin is, generally,
not conserved.  It was shown in \cite{isospin} that partial restoration of the isospin symmetry,
and validity of the $M^{2\nu}_{\mathrm{Fcl}} = 0$, can be achieved within the QRPA by choosing the
isospin symmetry for the $T = 1$ nucleon-nucleon interaction, i.e., by choosing the same strength
for the neutron-neutron and proton-proton pairing force treated within the BCS method, and
the isovector neutron-proton interaction 
treated by the QRPA equations of motion. (In practice,
the five effective coupling constants are close to each other,
but not exactly equal since the renormalization of 
the pairing strength
couplings $d_{nn}^{i,f}$ and $d_{pp}^{i.f}$ are
adjusted to reproduce the corresponding neutron and proton gaps and the neutron-proton
isovector coupling  renormalization $g_{pp}^{T=1}$ 
is chosen to reproduce the  $M^{2\nu}_{\mathrm{Fcl}} = 0$ relation.)  The values of these parameters are
shown in Table \ref{tab:1}.

$M^{2\nu}_{\mathrm{Fcl}} = 0$  follows from the isospin conservation and 
implies that $M_{S=0} = - M_{S=1}$. However,  both could be large in absolute value.
In Fig \ref{fig:s01} QRPA results for $M_{\mathrm{Fcl}}$=0 and $M_{\mathrm{GTcl}}$=0 are presented. In that case there is a 
significant difference in behavior of $C(r)$ for the $S=0$ and $S=1$, with the $S=0$ part
significantly larger that the $S=1$ part. 
 Note that in QRPA the values of   $M_{\mathrm{Fcl}}$=0 and $M_{\mathrm{GTcl}}$=0  depend on the already fixed 
renormalization strength $g_{pp}^{T=1}$
and on the value of $g_{pp}^{T=0}$. The $M_{S=0}$ values in Fig \ref{fig:s01}
are in agreement with the discussion in the preceding
section, where we saw that their values are numerically close to zero, actually oscillating between the positive
and negative values for different nuclei, and depending sensitively on the
properties of the poorly known higher lying $1^+$ states.

We can fulfil the relation  $M^{2\nu}_{\mathrm{GTcl}} = 0 $ by adjustment of the renormalization of the isoscalar
neutron-proton coupling strength $g^{T=0}_{pp}$. As we effectively restored the isospin symmetry by
the proper choice of $g^{T=1}_{pp}$, choosing $g^{T=0}_{pp}$ so that $M^{2\nu}_{\mathrm{GTcl}} = 0 $, corresponds
to the partial restoration of the spin-isospin symmetry $\mathrm{SU}(4)$. Obviously, choosing the effective neutron-proton
interaction in this way is quite different from the proposal in ref. \cite{Xav18} where the proportionality between
$M^{0\nu}_{\mathrm{GT}}$ and  $M^{2\nu}_{\mathrm{GTcl}}$ was proposed. We believe that assuming that  $M^{2\nu}_{\mathrm{GTcl}}
\sim 0$ reflects better the physics of the problem. Once the   $M^{2\nu}_{\mathrm{GTcl}}$ and   $M^{2\nu}_{\mathrm{Fcl}}$
have been fixed, the corresponding radial functions $C^{2\nu}_{\mathrm{GTcl}}(r)$ can be obtained, and from them, 
using Eq. (\ref{eq:c02}), the values of $M^{0\nu}_\mathrm{F}$ and $M^{0\nu}_{\mathrm{GT}}$ can be obtained. The results 
are described and discussed in the following section. 

Since we know the experimental values of the $2\nu\beta\beta$ matrix elements $M^{2\nu}$, it is legitimate
to ask whether the fact that they do not vanish can be compatible with our assumption that  the closure
matrix elements $M^{2\nu}_{\mathrm{GTcl}}$ vanish. Clearly, if $\bar{E}_{av}$ is the properly averaged energy denominator,
then
\begin{equation}
\bar{E} \times  M^{2\nu} = M^{2\nu}_{\mathrm{GTcl}}
\end{equation}
must be obeyed. If the right-hand side of this equation is vanishing, then one of the factors on the left-hand side
must vanish as well. In our case it must be the average energy  $\bar{E}$ reflecting the fact that in both
$M^{2\nu}$ and $M^{2\nu}_{\mathrm{GTcl}}$ are both positive and negative contributions to the corresponding sums
(by treating the negative sign in the numerator of (\ref{eq:2nme}) as negative denominator).

In our approach the parameter $g^{T=0}_{pp}$ is fixed by the 
requirement that $M^{2\nu}_{\mathrm{GTcl}} = 0$, it is thus straightforward
to evaluate, within QRPA, the $M^{2\nu}$ and compare them with their experimental values derived from the
observed $2\nu\beta\beta$ half-lives. In agreement with the idea of `$g_{\rm A}$ quenching', the calculated matrix
elements are typically larger than the experimental values. That discrepancy can be, at least in part, remedied
by choosing the effective $g_{\rm A}$ value, $g_{\rm A}^\mathrm{eff} = q \times g_{\rm A}^\mathrm{free}$. (Even somewhat better agreement
is achieved by assuming that $g_{\rm A}^\mathrm{eff}$ scales like $1/A^{1/2}$. We do not see any obvious justification
for such a dependence, and use $g_{\rm A}^\mathrm{eff}$ independent of A.) Taking the average ratio
of the calculated and experimental matrix elements, we arrive at $q = 0.712$. 
The resulting quenched
calculated matrix elements are compared with the experimental ones in Table \ref{tab:1}. The agreement
is only within a factor of $\sim2$, reflecting the known strong sensitivity of $M^{2\nu}$ on the 
$g^{T=0}_{pp}$ values.

\begin{table*}[!t]
  \caption{Renormalization parameters of the pairing interaction $d^{i,f}_{p,n}$ (i- initial nucleus, f-final nucleus, p-protons, n-neutrons)
    adjusted to reproduce experimental pairing gaps. Renormalization parameters of the isovector $g^{T=1}_{pp}$ and isoscalar $g^{T=0}_{pp}$
    particle-particle interactions of the residual Hamiltonian adjusted to reproduce, respectively, $M^{2\nu}_{\mathrm{Fcl}}=0$
    and $M^{2\nu}_{\mathrm{GTcl}}=0$ - an effective restoration of the isospin $\mathrm{SU}(2)$ and spin-isospin $\mathrm{SU}(4)$. The corresponding
    values of the $2\nu\beta\beta$-decay Fermi $M^{2\nu}_{\mathrm{F}}$ and  Gamow-Teller $M^{2\nu}_{\mathrm{GT}} \times q^2$
    matrix elements,
    where $q = 0.712$ is the effective quenching factor, $g_{\rm A}^{\rm eff} = q \times g_{\rm A}^{\rm free} = 0.904$. In the last column are
    the experimentally determined matrix elements $M^{2\nu}_\mathrm{exp}$ for unquenched $g_{\rm A}$.
    \label{tab:1}}
\centering   
\renewcommand{\arraystretch}{1.1}    
\begin{tabular}{lcccccccccccccc}\hline\hline
  Nucleus  & & $d^i_{pp}$ &  $d^f_{pp}$ & & $d^i_{nn}$ &  $d^f_{nn}$ & & $g^{T=1}_{pp}$ & $g^{T=0}_{pp}$ & & $M^{2\nu}_{F}$  & $M^{2\nu}_{\mathrm{GT}} \times q^2$  &   $M^{2\nu}_\mathrm{exp}$   \\   
  & &           &            & &           &            & &              &             &  &   [MeV$^{-1}$] &    [MeV$^{-1}$]  &   [MeV$^{-1}$]                    &   \\
 \hline
 $^{48}$Ca  & &     -     &    1.069   & &   -       &   0.982    & &     1.028    &   0.745     &  &   -0.003      &     0.019   & 0.046 &  \\
 $^{76}$Ge  & &    0.922  &    0.960   & &   1.053   &   1.085    & &     1.021    &   0.733     &  &    0.003      &     0.077   & 0.136  &  \\
 $^{82}$Se  & &    0.861  &    0.921   & &   1.063   &   1.108    & &     1.016    &   0.737     &  &    0.001      &     0.071   & 0.100 &  \\
 $^{96}$Zr  & &    0.910  &    0.984   & &   0.752   &   0.938    & &     0.961    &   0.739     &  &    0.001      &     0.162   & 0.097  &  \\
 $^{100}$Mo & &    1.000  &    1.021   & &   0.926   &   0.953    & &     0.985    &   0.799     &  &   -0.001      &     0.306   & 0.251  &  \\
 $^{116}$Cd & &    0.998  &     -      & &   0.934   &   0.890    & &     0.892    &   0.877     &  &   -0.000      &     0.059   & 0.136  &  \\
 $^{128}$Te & &    0.816  &    0.857   & &   0.889   &   0.918    & &     0.965    &   0.741     &  &    0.017      &     0.076   & 0.052  & \\
 $^{130}$Te & &    0.847  &    0.922   & &   0.971   &   1.011    & &     0.963    &   0.737     &  &    0.016      &     0.065   & 0.037 &   \\
 $^{136}$Xe & &    0.782  &    0.885   & &     -     &   0.926    & &     0.910    &   0.685     &  &    0.014      &     0.036   & 0.022 &  \\
  \hline\hline
\end{tabular}
\end{table*}

%---------------------------------------------------------
\section{$0\nu\beta\beta$ matrix elements and the partial $\mathrm{SU}(4)$ symmetry restoration.}
\label{S:SU}
%----------------------------------------------------------

\begin{figure}[htb]
%\centerline{\psfig{file=4thpaper_fig1.pdf,width=8.5cm}}
\includegraphics[width=\columnwidth]{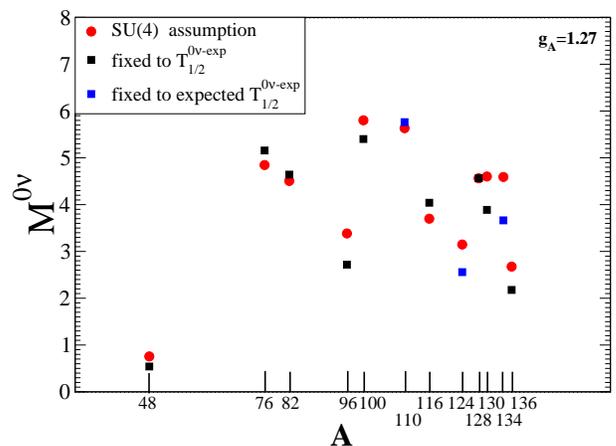}
\caption{$M^{0\nu}$ evaluated when $g_{pp}^{T=0}$ is adjusted so that the $2\nu$ half-life is correctly
reproduced (black squares) or by requiring that  $M^{2\nu}_{\mathrm{GTcl}} = 0$, i.e., partial restoration of the
$\mathrm{SU}(4)$ symmetry.}
\label{fig:su4_res}
\end{figure}

The matrix elements $M^{2\nu}$ of the $2\nu\beta\beta$ decay involve 
only $1^+$ virtual intermediate states. Within the QRPA they sensitively depend
on the magnitude of the isoscalar neutron-proton interaction \cite{martin}, conventionally denoted as 
$g_{pp}^{T=0}$. On the other hand,
matrix elements $M^{0\nu}$ of the $0\nu\beta\beta$ decay contain many multipoles of the intermediate
states.  Among them the $1^+$, or GT, is particularly sensitive to the $g_{pp}^{T=0}$; other multipoles are less
dependent to its magnitude.  That led to the practice \cite{us1,us2}, commonly used in QRPA now, to adjust
the $g_{pp}^{T=0}$ so that the experimental half-life $T^{2\nu}_{1/2}$ is correctly reproduced.  That way
the most sensitive multipole contributing to $M^{0\nu}$ has been tied to the experimentally determined 
quantity. (Also, it turns 
out that with this adjustment, the magnitude of $M^{0\nu}$ becomes essentially independent on the
size of the single particle basis included.) 

As explained above, in this work we propose instead to use the condition $M^{2\nu}_{\mathrm{GTcl}} = 0$,
i.e., partial restoration of the $\mathrm{SU}(4)$ symmetry, to adjust the value of the renormalization parameter
$g_{pp}^{T=0}$. The matrix elements $M^{0\nu}$ evaluated by these two alternative methods are
shown in Table \ref{table.2} together with the corresponding partial values  $M_\mathrm{F}$, $M_{\mathrm{GT}}$ and $M_\mathrm{T}$
separated into the spin $S = 0$ and $S = 1$ components. Few candidate nuclei ($^{94}$Zr, $^{110}$Pd,
$^{124}$Sn and $^{134}$Xe), where the $2\nu$ decay has not been observed as yet, are also included
in Table \ref{table.2}. All entries there were obtained when the sum over the virtual intermediate states
was explicitly evaluated. When the closure approximation is used together with the $\mathrm{SU}(4)$ adjustment, 
the results are similar, with the final $M^{0\nu}$ values about 10\% smaller, similar to the previous
experience described above.  Typically, the contributions of the spin $S = 1$ component
to the $M_\mathrm{F}$ and $M_{\mathrm{GT}}$ are indeed negligible. However, the tensor mart, $M_{\rm T}$ gets its value
only from $S = 1$; it constitutes about 10\% of the total $M^{0\nu}$ value.

Adjusting $g_{pp}^{T=0}$ to the condition of partial restoration of the $\mathrm{SU}(4)$ symmetry means that
the $2\nu$ matrix elements (and, naturally, the half-lives $T^{2\nu}_{1/2}$) are not any longer 
tied to their experimental values. The theoretical values of $M^{2\nu}$ are only in qualitative
agreement with experiment, as we saw in the previous section. However, remarkably, the new
adjustment of $g_{pp}^{T=0}$ causes only relatively small changes in the $M^{0\nu}$ as one could
see in Table \ref{table.2}. In Fig. \ref{fig:su4_res} the two ways of the $g_{pp}^{T=0}$ adjustment
are compared. The largest effect, for $^{130}$Te and $^{136}$Xe is an increase of $M^{0\nu}$ by
$\sim$ 20\%. Note that both variants shown in Fig. \ref{fig:su4_res} were evaluated with $g_{\rm A} = 1.27$,
i.e., without quenching.

\begin{table*}[!t]
\caption{The nuclear matrix elements (NME) associated with light neutrino mass mechanism of the $0\nu\beta\beta$-decay
    calculated within the proton-neutron QRPA  using two ways of fixing the strengths of residual interactions in the nuclear Hamiltonian:
    i) $g^{T=1}_{pp}$ and $g^{T=0}_{pp}$ are adjusted to reproduce $M^{2\nu}_\mathrm{F}=0$ and the experimental
$2\nu\beta\beta$ half-life,
  respectively ($T^{2\nu}_{1/2}$);
    ii) $g^{T=1}_{pp}$ and $g^{T=0}_{pp}$ are adjusted to reproduce $M^{2\nu}_{\mathrm{Fcl}}=0$
    and $M^{2\nu}_{\mathrm{GTcl}}=0$ - an effective restoration of the isospin $\mathrm{SU}(2)$ and spin-isospin $\mathrm{SU}(4)$
    symmetry. In i) and ii) the sum over all virtual excitations is explicitly performed. The partial Fermi, 
    Gamow-Teller, tensor and full $0\nu\beta\beta$-decay NME
    are presented for $S=0$ and $S=1$ channels and for the sum of them. Unquenched value of axial-vector coupling
    constant ($g_{\rm A}=1.269$),  Argonne two-nucleon short-range correlations and $\bar{E} = 8$ MeV are considered. 
    \label{table.2}}
\centering 
\renewcommand{\arraystretch}{1.1}    
\begin{tabular}{lccccccccccccccccc}\hline\hline
  Nucl.  & par. 
  & & \multicolumn{4}{c}{ S=0 }
  & & \multicolumn{4}{c}{ S=1 }
  & & \multicolumn{4}{c}{ full NME }  \\ \cline{4-7} \cline{9-12} \cline{14-17}
  &        
  & & $M_{\rm F}$ & $M_{\mathrm{GT}}$ & $M_{\rm T}$ & $M^{0\nu}$ 
  & & $M_{\rm F}$ & $M_{\mathrm{GT}}$ & $M_{\rm T}$ & $M^{0\nu}$
  & & $M_{\rm F}$ & $M_{\mathrm{GT}}$ & $M_{\rm T}$ & $M^{0\nu}$ & \\ \hline
  ${^{48}}$Ca & $T^{2\nu}_{1/2}$ &  & -0.253 & 0.659 & 0.00 & 0.816 & & -0.027 & -0.021 & -0.156 & -0.161  & & -0.280 & 0.638 & -0.156 & 0.656 &  \\
             & $\mathrm{SU}(4)$ & & -0.285 & 0.748 & 0.00 & 0.925 & & 0.006 & 0.009 & -0.158 & -0.153 & & -0.280 & 0.757 & -0.158 & 0.773 &  \\
  ${^{76}}$Ge & $T^{2\nu}_{1/2}$   & & -1.719 & 4.482 & 0.00 & 5.550 & & 0.111 & 0.102 & -0.588 &-0.554  & & -1.608 & 4.584 & -0.588 & 4.995 &  \\
             & $\mathrm{SU}(4)$  & & -1.705 & 4.443 & 0.00 & 5.502 & & 0.097 & 0.089 & -0.588 & -0.559 & & -1.570 & 4.455 & -0.583 & 4.846 &  \\
  ${^{82}}$Se & $T^{2\nu}_{1/2}$  & & -1.537 & 3.995 & 0.00 & 4.949 & & 0.037 & 0.035 & -0.544 & -0.532 & & -1.500 & 4.029 & -0.544 & 4.417 &  \\
             & $\mathrm{SU}(4)$ & & -1.587 & 4.133 & 0.00 & 5.119 & & 0.089 & 0.082 & -0.540 & -0.513 & & -1.499 & 4.216 & -0.540 & 4.606 &  \\
  ${^{94}}$Zr & $\mathrm{SU}(4)$ &  & -1.171 & 3.066 & 0.00 & 3.793 & & -0.066 & -0.050 & -0.392 & -0.401 & & -1.237 & 3.016 & -0.392 & 3.392 &  \\ 
  ${^{96}}$Zr  & $T^{2\nu}_{1/2}$ & & -0.916 & 2.359 & 0.00 & 2.928 & & -0.272 & -0.242  & -0.420 & -0.494 & & -1.188 & 2.117 & -0.420 & 2.435 &  \\
              & $\mathrm{SU}(4)$ &  & -1.174 & 3.069 & 0.00 & 3.798 & & -0.008 & -0.001 & -0.405 & -0.401 & & -1.182 & 3.068 & -0.405 & 3.396 &  \\
  ${^{100}}$Mo & $T^{2\nu}_{1/2}$ & & -1,799 & 4.658 & 0.00 & 5.775 & & -0.410 & -0.362 & -0.707 & -0.814 & & -2.209 & 4.296 & -0.707 & 4.961 &  \\
              & $\mathrm{SU}(4)$ &  & -2.038 & 5.327 & 0.00 & 6.592 & & -0.168 & -0.136 & -0.692 & -0.724 & & -2.206 & 5.191 & -0.692 & 5.868 &  \\
  ${^{110}}$Pd & $\mathrm{SU}(4)$ &  & -1.961 & 5.115 & 0.00 & 6.332 & & -0.174 & -0.145 & -0.607 & -0.643 & & -2.135 & 4.970 & -0.607 & 5.689 &  \\
  ${^{116}}$Cd & $T^{2\nu}_{1/2}$ & & -1.280 & 3.328 & 0.00 & 4.123 & & 0.274 & -0.235 & -0.290 & -0.355 & & -1.554 & 3.093 & -0.290 & 3.768 &  \\
              & $\mathrm{SU}(4)$ &  & -1.272 & 3.305 & 0.00 & 4.095 & & -0.283 & -0.243 & -0.291 & -0.358 & & -1.555 & 3.062 & -0.291 & 3.737 &  \\
  ${^{124}}$Sn & $\mathrm{SU}(4)$ &  & -1.096 & 2.862 & 0.00 & 3.543 & & 0.032 & 0.031 & -0.347 & -0.336 & & -1.064 & 2.894 & -0.347 & 3.207 &  \\
   ${^{128}}$Te & $T^{2\nu}_{1/2}$ &  &-1.638 & 4.248 & 0.00 & 5.265 & & -0.146 & -0.125 & -0.604 & -0.638 & & -1.784 & 4.122 & -0.604 & 4.626 &  \\
              & $\mathrm{SU}(4)$ &  & -1.839 & 4.784 & 0.00 & 5.923 & & -0.044 & -0.033 & -0.588 & -0.594 & & -1.878 & 4.751 & -0.588 & 5.329 &  \\
  ${^{130}}$Te & $T^{2\nu}_{1/2}$ &  & -1.411 & 3.655 & 0.00 & 4.531 & & -0.162 & -0.140 & -0.554 & -0.593 & & -1.573 & 3.515 & -0.554 & 3.939 &  \\
              & $\mathrm{SU}(4)$   &  & -1.616 & 4.215 & 0.00 & 5.219 & & -0.053 & -0.042 & -0.536 & -0.545 & & -1.669 & 4.173 & -0.536 & 4.673 &  \\
  ${^{134}}$Xe & $\mathrm{SU}(4)$   &  & -1.598 & 4.163 & 0.00 & 5.156 & & -0.044 & -0.034 & -0.498 & -0.504 & & -1.642 & 4.129 & -0.498 & 4.652 &  \\
  ${^{136}}$Xe & $T^{2\nu}_{1/2}$ &  & -0.780 & 2.009 & 0.00 & 2.493 & & -0.035 & -0.028 & -0.285 & -0.291 & & -0.815 & 1.980 & -0.285 & 2.202 &  \\
              & $\mathrm{SU}(4)$   &  & -0.927 & 2.410 & 0.00 & 2.985 & &  0.022 &  0.022 & -0.274 & -0.266 & & -0.905 & 2.432 & -0.274 & 2.720 &  \\
  \hline \hline
\end{tabular}
\end{table*}

%---------------------------------------
\section{Summary}
\label{S:Sum}
%---------------------------------------

In this work we discuss the importance of dependence of the $0\nu$ and $2\nu$ nuclear matrix elements
on the distance $r_{ij}$ between the two neutrons that are transformed in two protons in the double-beta
decay. We show that, if this function, $C(r)$, is known for any particular mechanism of the decay, evaluation
of the matrix element for any other mechanism is reduced to an integral using Eq. (\ref{eq:c02}).

Further, we show that there is a close relation between the GT part of the $M^{0\nu}$ and the
matrix element of the experimentally observed $2\nu\beta\beta$ decay, evaluated however in the
closure approximation, $M^{2\nu}_{\mathrm{cl}}$. Our work does not support the conjecture in Ref.\cite{Xav18}
of proportionality between the $M^{0\nu}_{\mathrm{GT}}$ and $M^{2\nu}_{\mathrm{cl}}$. Instead, we argue that the positive
contributions to $M^{2\nu}_{\mathrm{cl}}$ from the lower lying $1^+$ intermediate states is essentially fully 
cancelled by the negative contribution of the higher lying $1^+$ states. We also show that the
contribution of the triplet spin $S = 1$ two neutron states is much smaller than the contribution of the
singlet $S = 0$ states. (Note that  when $M^{2\nu}_{\rm F}$ = 0 the $S=0$ part is always
three times larger that the $S=1$ part.) From these considerations follows a simple proportionality 
between  the Fermi and GT parts of the $M^{2\nu}_{\mathrm{cl}}$. 

Based on these consideration we arrive at a new way of adjusting the important QRPA parameter, the 
renormalization of the isoscalar particle-particle interaction, $g_{pp}^{T=0}$. We propose that
its value should be determined
from the requirement that $M^{2\nu}_{\mathrm{GTcl}} = 0$. Together with $M^{2\nu}_{\mathrm{Fcl}} = 0$, following
from isospin conservation, these two condition are equivalent to the restoration of partial conservation
of spin-isospin $\mathrm{SU}(4)$ symmetry.

We then evaluate the true $2\nu$ matrix elements and compare them to the corresponding experimental 
values. The calculated $M^{2\nu}$ values are mostly larger than the experimental ones, suggesting 
on average a relatively modest quenching $g_{\rm A}^\mathrm{eff} = 0.710 \times g_{\rm A}^\mathrm{free}$. The agreement
between the calculated and experimental values of $M^{2\nu}$ is, however, only qualitative. That is, perhaps,
not surprising given the strong dependence of the calculated $M^{2\nu}$ values on the $g_{pp}^{T=0}$.

The $0\nu$ matrix elements, corresponding to the ``standard" light Majorana neutrino exchange are
evaluated next using the new adjustment of the $g_{pp}^{T=0}$. When they are compared to
the values obtained when $g_{pp}^{T=0}$ is chosen so that the $2\nu$ half-life is correctly reproduced,
which was a QRPA standard procedure until now,
only relatively modest changes of the $M^{0\nu}$ are obtained. This shows that, within QRPA, the $M^{0\nu}$ values 
are quite stable. It also, in our opinion, represents a better way to determine the 
parameter $g_{pp}^{T=0}$, and through the corresponding function $C^{2\nu}_{\mathrm{GTcl}}(r)$ all 
possible $0\nu$ nuclear matrix elements.  

%------------------------------------------
\section{Acknowledgment}
%-------------------------------------------
This work was supported by the VEGA Grant Agency of the Slovak Republic under Contract No. 1/0922/16, by Slovak Research and Development Agency under Contract No. APVV-14-0524, RFBR Grant No. 16-02-01104, Underground laboratory LSM - Czech participation to European-level research infrastructure CZ.02.1.01/0.0/0.0/16 013/0001733. The work of P.V. is supported by the Physics Department, California Institute of Technology.

%----------------------------------------
\section{Appendix: $LS$ coupling scheme}
\label{S:App}

In the QRPA the closure matrix element ${M}^{\alpha}_{K}$
($\alpha = 0\nu,~2\nu$ and $K=\mathrm{F}$ (Fermi), $\mathrm{GT}$ (Gamow-Teller)
and $\mathrm{T}$ (Tensor)) can be written as
a sum over two neutron (initial nucleus) and two proton (final nucleus) states
participating in the two virtual beta decays inside nucleus, angular momentum $\mathcal{J}$
to which they are coupled, and angular momentum and parity $J^{\pi}$ of the
intermediate nucleus as follows:
\begin{eqnarray}
\label{eq:long}
M^\alpha_{K} & = & \sum_{pnp'n'} \sum_{J^{\pi}\mathcal{J}} 
(-1)^{j_n + j_{p'} + J + {\mathcal J}} \sqrt{ 2 {\mathcal J} + 1} \times\qquad\qquad
\\
&&
\left\{
\begin{array}{c c c}
j_p & j_n & J  \\
 j_{n'} & j_{p'} & {\mathcal J}
\end{array} 
\right\} ~D(p' n', p n; J^\pi)~
T^\alpha_K(p p', n n'; {\mathcal J}), \nonumber
\end{eqnarray}
where
\begin{eqnarray}
D(p' n', p n, J^\pi) & = &  \sum_{J^\pi,k_i,k_f} 
\langle 0_f^+ \parallel [ \widetilde{c_{p'}^+ \tilde{c}_{n'}}]_J \parallel J^{\pi} k_f \rangle \times
\\        
&& \langle  J^{\pi} k_f |  J^{\pi} k_i \rangle
 \langle  J^{\pi} k_i\parallel [c_p^+ \tilde{c}_n]_J \parallel 0_i^+ \rangle\, \nonumber
\end{eqnarray}
includes products of reduced matrix elements of one-body densities
$c_p^+ \tilde{c}_n$ ($\tilde{c}_n$ denotes the time-reversed state)
connecting the initial nuclear ground state with the final nuclear ground state through
a complete set of states of the intermediate nucleus labeled by their angular
momentum and parity, $J^{\pi}$, and indices $k_i$ and $k_f$. They depend 
on the BCS coefficients $u_i,v_j$ and on the QRPA vectors $X,Y$ \cite{isospin}.
The coupling $(lsj)$ for each single proton (neutron) state is considered,
i.e., the individual orbital momentum $l_p$ ($l_n$) and spin $s_p$ ($s_n$)
is coupled to the total angular momentum $j_p$ ($j_n$).
The non-antisymmetrized two-nucleon matrix element takes the form
\begin{eqnarray}\label{twobody}
T^\alpha_K(p p', n n'; {\mathcal J}) = \langle p(1) p'(2); {\mathcal J} \parallel 
{\cal O}^\alpha_K \parallel n(1) n'(2); {\mathcal J} \rangle 
\nonumber \\
\end{eqnarray}
where
\begin{eqnarray}
  &&  {\cal O}^{2\nu}_{\rm F} = 1,~~~~{\cal O}^{2\nu}_{\mathrm{GT}} = \sigma_{12},~~~~{\cal O}^{2\nu}_{\rm T} = S_{12} \nonumber\\
  &&  {\cal O}^{0\nu}_{\mathrm{F,GT,T}}(r_{12})  = {\cal O}^{2\nu}_{\mathrm{F,GT,T}} ~ H_K(r_{12}, \bar{E})
\end{eqnarray}
with $S_{12} = 3 (\vec{\sigma}_1 \cdot \hat{r}_{12}) (\vec{\sigma}_2 \cdot \hat{r}_{12}) - \sigma_{12}$,
$\sigma_{12} = \vec{\sigma}_1 \cdot \vec{\sigma}_2$. $\vec{r}_{12} = \vec{r}_1 - \vec{r}_2$, 
${r}_{12} = |\vec{r}_{12}|$ and  $\hat{r}_{12} = \vec{r}_{12}/r_{12}$, where $\vec{r}_1$ and 
$\vec{r}_2$ are coordinates of nucleons undergoing beta decay. 
For the exchange of light Majorana neutrinos, the $0\nu\beta\beta$ decay mechanism we
are considering here, the neutrino potentials $H_K (r_{12}, \bar{E})$ are given in Eq. (\ref{HGT})

It practice, the calculation of non-antisymmetrized two-nucleon matrix
element in Eq. (\ref{twobody}) is performed in center of mass frame
by using a harmonic oscillator single particle basis set. The transformation
from $jj$ to $LS$ coupling is used and 
the Talmi transformation via the Moshinsky transformation brackets
is considered. In the case of the $0\nu\beta\beta$-decay two-nucleon
matrix elements we obtain 
\begin{eqnarray}
  && \left(
\begin{array}{l}
  T^{0\nu}_{\rm F}   \\
  T^{0\nu}_{\mathrm{GT}} \\
    T^{0\nu}_{\rm T} 
\end{array}\right)(p p', n n'; {\mathcal J})
=  \hat{\mathcal J} \hat{j}_{n} \hat{j}_{n'} \hat{j}_{p} \hat{j}_{p'}
\sum\limits_{SL} (2S+1)\times \nonumber \\
&& ~~~~~(2L+1)\left\{
\begin{array}{lll}
1/2 & l_p   & j_p  \\
1/2 & l_{p'} & j_{p'}\\
S & L & {\mathcal J}
\end{array}
\right\}
\left\{
\begin{array}{lll}
1/2 & l_n   & j_n  \\
1/2 & l_{n'} & j_{n'}\\
S & L & {\mathcal J}
\end{array}
\right\}\times \nonumber\\
&&\sum\limits_{n l n' l' \atop{\cal N L}}
\langle n l, {\cal N L}, L | n_p l_p, n_{p'} l_{p'}, L \rangle \times\nonumber\\
&&~~~~~~~\langle n' l', {\cal N L}, L | n_n l_n, n_{n'} l_{n'}, L \rangle
\sum\limits_{J'} (2J'+1) \times \nonumber\\
&& ~~~~~~\sqrt{(2l+1)(2l'+1)}
\left\{
\begin{array}{l@{\,}l@{\,}l}
l & L         & {\cal L}\\
{\cal J} & J'  & S
\end{array}
\right\}
\left\{
\begin{array}{l@{\,}l@{\,}l}
l' & L       & {\cal L}\\
{\cal J} & J'  & S
\end{array}
\right\}
\times\nonumber\\
&&~~
\langle n l, S; J' |
\left(
\begin{array}{c}
  (~~\delta_{S 0} + \delta_{S 1})H_{\rm F}(r_{12},\bar{E})\\
  (-3\delta_{S 0} + \delta_{S 1})H_{\mathrm{GT}}(r_{12},\bar{E})\\
   S_{12}H_{\rm T}(r_{12}.\bar{E})
\end{array}\right)
| n' l', S; J' \rangle.
\nonumber\\&&
\end{eqnarray}
Here, $\hat{\cal J} = \sqrt{2 {\cal J} +1}$ 
and $\hat{j}_\alpha = \sqrt{2 j_\alpha +1}$ with $\alpha = p,~p',~n$ and $n'$ .
We note that in the case of the Fermi and Gamow-Teller transitions there are both
$S=0$ an $S=1$ contributions, unlike the case of the tensor transition where
only $S=1$ is allowed. Due to the presence of  neutrino potentials
$H_K (r_{12},\bar{E})$  ($K = \mathrm{F}$, $\mathrm{GT}$ and $\mathrm{T}$) in two-body transition operators
there is dominance of the $S=0$ contribution
to $M^{0\nu}$. There is a small
difference between the Fermi and Gamow-Teller neutrino potentials due
to a different form factor's cut-off and contributions from higher
order terms of the nucleon currents. If they would be equal, and
the $S=1$ contribution could be neglected, we would end up with 
\begin{eqnarray}
M^{0\nu}_{\mathrm{GT}} \simeq -3  M^{0\nu}_{\mathrm{F}}.
\end{eqnarray}

The $2\nu\beta\beta$-decay Fermi and Gamow-Teller matrix elements can be
decomposed into the $S=0$ and $S=1$ contributions as follows
(see Eq. (\ref{eq:sepr})):
\begin{eqnarray}
M^{2\nu}_{\mathrm{GT}} &=& -3 M^{2\nu}_{S=0}  + M^{2\nu}_{S=1}, \nonumber\\ 
M^{2\nu}_{\mathrm{F}} &=&  ~~~~  M^{2\nu}_{S=0}  + M^{2\nu}_{S=1} 
\end{eqnarray}  
The corresponding decomposition of the non-antisymmetrized two-nucleon
matrix element is given by
\begin{eqnarray}
  && \left(
\begin{array}{l}
  T^{2\nu}_{\rm F}   \\
  T^{2\nu}_{\mathrm{GT}} 
\end{array}\right)(p p', n n'; {\mathcal J})
=  \hat{\mathcal J} \hat{j}_{n} \hat{j}_{n'} \hat{j}_{p} \hat{j}_{p'}
\sum\limits_{SL} (2S+1)\times \nonumber \\
&& ~~~~~(2L+1)
\left\{
\begin{array}{lll}
1/2 & l_p   & j_p  \\
1/2 & l_{p'} & j_{p'}\\
S & L & {\mathcal J}
\end{array}
\right\}
\left\{
\begin{array}{lll}
1/2 & l_n   & j_n \\
1/2 & l_{n'} & j_{n'}\\
S & L & {\mathcal J}
\end{array}
\right\}\times \nonumber\\
&& ~~~~~\delta_{n_{p} n_{p'}} \delta_{l_{p} l_{p'}} \delta_{n_{n} n_{n'}} \delta_{l_{n} l_{n'}} \times
\left(
\begin{array}{c}
 \delta_{S 0} + \delta_{S 1} \\
 -3 \delta_{S 0} + \delta_{S 1} \\
\end{array}\right)
\end{eqnarray}
If $M^{2\nu}_{\mathrm{F}}=0$ because of  isospin conservation
(see \cite{isospin}), then $S=0$ and $S=1$ contributions are equal in magnitude
but opposite in sign.

%----------------------------------------------------

% Bibliography


\begin{thebibliography}{99}

\bibitem{SV} J. Schechter and J.W.F. Valle, Phys.Rev.D {\bf 25}, 2951 (1982).

\bibitem{EM} J. Engel and J. Menendez, Rept. Prog. Phys. {\bf 80}, 046301 (2017).

\bibitem{SHFV} F. \v{S}imkovic, R. Hod\'{a}k, A. Faessler, and P. Vogel,
   Phys. Rev. C {\bf 83}, 015502 (2011).

\bibitem{PDG17} C. Patrignani et al. [Particle Data Group], Chin. Phys. C {\bf 40}, 100001 (2016) and update 2017.

\bibitem{anatomy} F. \v{S}imkovic, A. Faessler, V. A. Rodin, P. Vogel, and J. Engel, Phys. Rev. C {\bf 77}, 045503 (2008).

\bibitem{SPVF}  F. \v{S}imkovic, G. Pantis, J. D. Vergados, and A. Faessler, Phys. Rev. C {\bf 60}, 055502 (1999).

\bibitem{Sim09} F. \v{S}imkovic,  A. Faessler, H. Muther, V. Rodin, and M. Stauf, Phys. Rev. C {\bf 79}, 055501 (2009).

\bibitem{KI12} J. Kotila and F. Iachello, Phys. Rev. C {\bf 85}, 034316 (2012).

\bibitem{HN17} M. Horoi and A. Neacsu, arXiv:1706.05391[hep-ph].

\bibitem{SH} R. A. Sen'kov and M. Horoi, Phys. Rev. C {\bf 90}, 051301 (2014).

\bibitem{XM9} J. Menendez {\it et al.} Nucl. Phys. A {\bf 818}, 130 (2009).

\bibitem{P17} S. Pastore {\it et al.}, Phys. Rev. C {\bf 97}, 014606 (2018).

\bibitem{Horoi07} M. Horoi, S. Stoica and B.A. Brown, Phys.Rev.C {\bf 75}, 034303 (2007).

\bibitem{SSD} J. Abad, A. Morales, R. Nunez-Lagos, and A.F. Pacheco, An. Fis. A {\bf 80}, 9 (1984);
  J. Phys. (Paris) {\bf 45}, 147 (1984).

\bibitem{Grewe} E.W. Grewe {\it et al}., Phys. Rev. C {\bf 78}, 044301 (2008).

\bibitem{coello} E.A. Coello-Perez, J. Menendez and A. Schwenk, arXiv:1708.06140[nucl-th].

\bibitem{ejiri12} H. Ejiri, J. Phys. Soc. Japan {\bf 81}, 033201 (2012).

\bibitem{SSD1} P. Domin, S. Kovalenko, F. \v{S}imkovic and S.V. Semenov, Nucl. Phys. {\bf A753}, 337 (2005).

\bibitem{SSD3} R. Arnold {\it et al.}, arXiv:1806.05553[hep-ex].
  
\bibitem{SSD2} F. \v{S}imkovic, R. Dvornick\'y, D. \v{S}tef\'anik and A. Faessler, Phys. Rev. C {\bf 97}, 034315 (2018).

\bibitem{Xav18} N. Shimizu, J. Menendez and K. Yako, Phys.Rev.Lett.{\bf 120}, 142502 (2018).

\bibitem{isospin} F. \v{S}imkovic, V. A. Rodin, A. Faessler, and P. Vogel,  Phys. Rev. C {\bf 87}, 045501(2013).

\bibitem{martin} P. Vogel and M.R. Zirnbauer, Phys. Rev. Lett.{\bf  57}, 3148 (1986).

\bibitem{us1} V. A. Rodin, A. Faessler, F. \v{S}imkovic and P. Vogel, Phys. Rev. C {\bf 68}, 044302(2003).

\bibitem{us2} V. A. Rodin, A. Faessler, F. \v{S}imkovic and P. Vogel,
        Nucl. Phys. {\bf A766}, 107 (2006), and erratum {\bf A793}, 213 (2007).



\end{thebibliography}
\end{document}